\documentstyle[psfig]{aanew}

\begin{document}

%
\title{\emph{Beppo}SAX view of NGC 526A: a Seyfert 1.9 galaxy with
a flat spectrum}

\author{R.~Landi\inst{1,2},  L.~Bassani\inst{2}, G.~Malaguti\inst{2},
M.~Cappi\inst{2}, A.~Comastri\inst{3}, M.~Dadina\inst{1}, 
           G.~Di~Cocco\inst{2}, A.C.~Fabian\inst{4}, E.~Palazzi\inst{2},
G.G.C.~Palumbo\inst{5}, M.~Trifoglio\inst{1}}

\offprints{R.~ Landi (landi@tesre.bo.cnr.it)}

\institute{Dipartimento di Fisica, Universit\`a di
             Bologna, via Irnerio 46 I--40127 Bologna \and
             ITeSRE/CNR, via Piero Gobetti 101, I--40129 Bologna,
             Italy \and 
              Osservatorio Astronomico di Bologna, via Ranzani 1, I--40127
              Bologna, Italy \and
              Institute of Astronomy, Madingley Road, Cambridge CB3 0HA, UK \and
              Dipartimento di Astronomia, Universit\`a di Bologna, via Ranzani 1,
              I--40127 Bologna, Italy}
\date{Received / accepted}

\abstract{
In the present work we report the \emph{Beppo}SAX observation of the
Seyfert 1.9 galaxy NGC 526A in the band 0.1--150 keV.
The high energy instrument onboard, PDS, has succeeded in measuring for
the first time the spectrum of this source in the 13--150 keV range.
The combined analysis of all Narrow Field Instruments 
provides a power law spectral index of $\simeq$ 1.6 and confirms
the flat spectral nature of this source.
Although NGC 526A varies strongly in the 2--10 keV over period of
months/years, its spectral shape
remains constant over these timescales.   
An Fe$_{\rm K \alpha}$ line, characterized by a complex structure, has
been 
detected in the 6--7 keV range. The line, which has an equivalent width
of 120 eV, is not compatible with being
produced in an absorbing medium with $N_{\rm H}\sim10^{22}$
cm$^{-2}$, but most likely originates by reflection in 
an accretion disk viewed at
an intermediate inclination angle of $\sim$ 42$^{\circ}$. The reflection
component is however small ($R\leq0.7$) and so it is not sufficient
to
steepen the spectrum to photon index values more typical of AGNs.
Instead, we find that the data are more consistent with
a flat power law spectrum cut--off at around 100 keV plus a small 
reflection component which could explain the observed iron line.
Thus NGC 526A is the only bona--fide Seyfert 2 galaxy which maintains a 
"flat spectrum" even when broad band data are considered: in this sense
its properties, with respect to the general class of Seyfert 2's, are
analogous to those of NGC 4151 with respect to the vast majority of
Seyfert 1's. 
\keywords{X--rays: galaxies --
                Galaxies: Seyfert --
                Galaxies: individual: NGC 526A
               }}

\authorrunning{R. ~Landi  et al.}
\titlerunning{\emph{Beppo}SAX view of NGC 526A
}

\maketitle

%

\section{Introduction}

NGC 526A, a prototype Narrow Emission Line Galaxy (NELG),
is a relatively nearby ($z=0.0192$) S0 peculiar type galaxy which is the
brightest member of a strongly interacting pair.
Although the source is often classified as a Seyfert
1--1.5 in various databases (see NED and Simbad), it is a true
example of a type 1.9 Seyfert.
In fact in NGC 526A discovery paper (Griffiths et al. 1979), it was stated
that
a broad component was only visible for H$\alpha$. This was subsequently
confirmed by Winkler (1992) who also reports lack of any broad H$\beta$
component during a more intense state of the broad H$\alpha$ line.
The low value of the narrow H$\alpha$/H$\beta$ ratio (=3.0) indicates that
the narrow line region is essentially unreddened, while the broad line
region suffers significant obscuration as the broad 
H$\alpha$/H$\beta$$\geq10$.\\ 
The galaxy, first
identified as an X--ray source by the \emph{HEAO--1}/SMC experiment
(Griffiths et al. 1979),
is one of the brightest extragalactic objects, being one of the
Piccinotti sample sources. As such it has been the target of various
X--ray
observations in both
soft and hard X--rays since its discovery (see Polletta et al. 1996 for a
recent compendium of X--ray data).

At soft energies, \emph{ROSAT} data are compatible with the source
being point--like and having a 0.1--2.4 keV flux of $0.2\times10^{-11}$
erg cm$^{-2}$ s$^{-1}$ (Rush et al. 1996). 
The source is strongly variable in the 2--10 keV band by a factor of
$\sim$~5.
The spectrum in this band is fairly simple being characterized by an 
absorbed ($N_{\rm H}\sim 2\times10^{22}$ cm$^{-2}$) power
law with a photon index of 1.5--1.6 (Smith $\&$ Done 1996; Turner et al.
1997):
since in NGC 526A the spectrum remains flat notwithstanding the use
of more complex models, the source has been taken as a prototype of the 
"flat spectrum" Seyfert 2.

This flat shape is incompatible with current
theories of Seyfert
unification schemes, which predict steeper intrinsic power law of
$\Gamma$~=~1.8--2.0 as generally
observed in Seyfert 1 galaxies. However,
discrepancies from these schemes have been reported for Seyfert 2 galaxies
in the sense that both flatter and steeper power law indices have been
observed (Smith $\&$ Done 1996; Awaki et al. 1991). In a few cases the
observations
have been reconciled with the Unification Scenario by introducing complex
absorption in the source  and/or reflection (Cappi et al. 1996; Vignali et
al. 1998; Malizia et al. 1999; Lanzi 2000; 
Malaguti et al. 1999). Furthermore, recent work using \emph{OSSE} data
alone or combined with low energy data, indicates that the average
intrinsic spectrum of
Seyfert 2 galaxies may be substantially harder than that of Seyfert 1,
again
in contrast with unified models of AGN (Zdziarski et al. 2000 and
references therein). Thus the problem of "flat spectrum" Seyfert 2
galaxies
is still an open issue, particularly for the relevance it may have with
respect to the synthesis of the X--ray background. 

The X--ray spectrum of NGC 526A is also characterized by a strong 
Fe$_{\rm K \alpha}$ line, most
likely due to an accretion disk: the data can be
modeled equally well with the superposition of a component from an
accretion disk viewed at 34$^\circ$  (Weaver $\&$ Reynolds 1998) plus a
torus contribution or with a 
single feature from a  nearly pole--on accretion disk (Turner et al.
1998). The observation of this line feature is in contrast with the 
lack of a strong reflection component: \emph{Ginga} data put an upper
limit to
$R$ (the strength of the reflected signal relative to the level of the
incident radiation) at 0.5 (Smith $\&$ Done 1996).  

NGC 526A has also been observed at high energies both by \emph{OSSE} and
\emph{BATSE}
on the \emph{CGRO}: the 20--100 keV flux was observed to be in the range
(8--12)~$\times~10^{-11}$ erg cm$^{-2}$ s$^{-1}$
(Malizia et al. 1999; Zdziarski et al. 2000).
Here we present the broad band (0.1--150 keV) spectrum of NGC 526A
obtained
by \emph{Beppo}SAX NFI, which defines the source spectral components and
consequently solves some of the still open
questions regarding this object. In particular, we demonstrate that the
source spectrum is indeed  flat as previously suggested and has an
exponential cut--off at around 100 keV. 
\section{Observation and data reduction} 
The \emph{Beppo}SAX X--ray observatory (Boella et al. 1997a) is a major
programme
of the Italian Space Agency with participation of the Netherlands Agency for 
Aereospace Programs.
 
This work concerns results obtained with three of the 
Narrow Field Instruments (NFI) onboard:
the Low Energy Concentrator Spectrometer (LECS; Parmar et al. 1997), 
the Medium Energy Concentrator Spectrometers 
(MECS; Boella et al. 1997b) and the Phoswich 
Detector System (PDS; Frontera et al. 1997).

\emph{Beppo}SAX NFI pointed at NGC 526A from Dec 31th 1998, to Jan 2th
1999.
The effective exposure times were $3.83\times10^4$ s for the LECS,
$9.32\times10^4$ s for the MECS and $4.64\times10^4$ s for the PDS.

The data reduction and analysis have been performed using {\sc FTOOLS 4.1}
and
{\sc XANADU 10.0} software packages using standard techniques and
selection criteria.

Another source (named 1SAX J0122.8--3448) has been detected in the MECS 
field of view at position 
$\alpha(2000)=01^{h}22^{m}47.5^{s}$,
$\delta$(2000)~=~--34$^{\circ}48'22^{\prime \prime}$, with an 
uncertain circle of 1$'$ in radius (more details in the Appendix).
  
Apart from NGC 526A and the new MECS detection, no other known 
sources emitting in the 2--150 band are known to be present in the PDS
field of view of 1.3 degrees (FWHM).
Due to the large difference in flux, it is unlikely that this new
MECS source contaminates the high energy emission of NGC 526A.

Light curves and spectra 
were extracted from a region centered in NGC 526A
with a radius of 4$'$ and 8$'$ for the MECS and the LECS respectively.
The source plus background light curves of each NFI did not 
indicate any significant variability;  
therefore, the data were integrated over the entire observing period to
obtain an average spectrum of the source.
LECS and MECS background subtraction was performed by means of blank sky 
deep field exposures accumulated by the \emph{Beppo}SAX Science Data
Center in the first year of the mission.
 
The PDS reduction was performed using the {\sc XAS}
software package and taking into account both rise time and spike corrections; 
products were obtained by plain subtraction of the "OFF" from the "ON"
source data.

No HPGSPC data will be considered in the present paper as the source is
too faint for a correct background subtraction.
\section{Spectral analysis}
LECS and MECS data were rebinned in order to sample the energy resolution of
the detector with an accuracy proportional to the count rate: two channels
for the LECS and four channels for the MECS. The PDS data were instead
rebinned so as to have logarithmically equal energy intervals. The
data
rebinning is also required to have at least 20 counts per channel to
ensure applicability of the $\chi^{2}$ statistics.
Spectral data from LECS, MECS  and PDS data  have 
been fitted simultaneously. We restricted the spectral analysis to 0.1--4.0
keV and 1.8--10.5 keV energy bands for the LECS and MECS 
respectively as these are
the energy ranges where the response matrices released in September 1997 are best 
calibrated. The PDS analysis was restricted to data below 150 keV as the
source signal disappears above this energy.

Normalization constants have been introduced to allow for known
differences in the absolute
cross--calibration between the detectors.

The values of the two constants have been allowed to vary.
Using the absorbed power law model the LECS/MECS and PDS/MECS constants 
turned out to be $0.62\pm0.04$ and $0.82\pm0.13$ respectively.
In all other models both constants have been constrain to lie within the
suggested intervals (Fiore, Guainazzi $\&$ Grandi 1999).

The spectral analysis has been performed by means of the {\sc XSPEC 10.0} 
package,
and using the instrument response matrices released by the \emph{Beppo}SAX
Science Data Centre in September 1997.
 
All the quoted errors correspond to 90$\%$ confidence intervals for one interesting 
parameter ($\Delta\chi^2=2.71$).
All the models used in what follows, contain an additional term to allow 
for the absorption of X--rays due to our galaxy, which in the direction of 
NGC 526A is caused by a column density of $2.2\times10^{20}$
cm$^{-2}$.

During the \emph{Beppo}SAX observation, NGC 526A
was at a 2--10 (20--100) keV flux level of $1.8~(2.4)\times10^{-11}$
erg cm$^{-2}$ s$^{-1}$, i.e. 
almost a factor of 2 (3--5) dimmer than when observed by \emph{ASCA
(BATSE/OSSE)} 
(Turner et al. 1997; Malizia et al. 1999; Zdziarski et al. 2000) 
but 60--70 $\%$ brighter than when measured 
by \emph{Ginga} (Smith $\&$ Done 1996).
\subsection{The 0.1-10.5 keV spectrum}
To check consistency with previous X--ray data, we first concentrated on 
the 0.1--10 keV band and fit the LECS--MECS spectrum with a simple 
absorbed power law model.
The fit is not satisfactory
($\chi^{2}/\nu=156/107$), as a soft excess and a line feature around 
6--7 keV are 
evident in the model to data ratio (Fig. 1), and results in a flat
spectrum having  
a photon index $\Gamma=1.45\pm{0.05}$, absorbed by a column density of 
$N_{\rm H}=(1.33\pm{0.15})\times10^{22}$ cm$^{-2}$.

The photon index is 
significantly lower than the canonical value generally found in Seyfert 
galaxies, confirming the flat spectral nature of NGC 526A reported by
\emph{Ginga} and \emph{ASCA} (Smith $\&$ Done  1996; Turner et
al. 1997); note that this
flat spectrum is found at different flux values,
indicating that it is not a result of spectral variability.
The absorbing column is consistent with previous measurements which
implies that the state of the absorber is not effected by the state
of the source.

In order to improve the quality of the fit, we fitted the data 
by adding to the previous model a narrow ($\sigma=0$ keV) gaussian line 
to account for excess emission around 6--7 keV.
The line feature is centered at $6.52^{+0.13}_{-0.16}$ keV
and has an equivalent width $EW=121^{+47}_{-44}$ eV.

If the line width is 
allowed to vary, the additional parameter gives a statistical
improvement in the fit ($\Delta\chi^{2} =2.6$ for one
additional parameter) and the width observed 
is $ 0.28^{+0.21}_{-0.16}$ keV. However, from the confidence contours of 
the
line against the energy, we conclude that the width is consistent with zero and
quote a 90$\%$ upper limit of 0.6 keV.
Similar values for the line parameters were also found by 
\emph{Ginga} and \emph{ASCA}.

%
%
\begin{figure}
\psfig{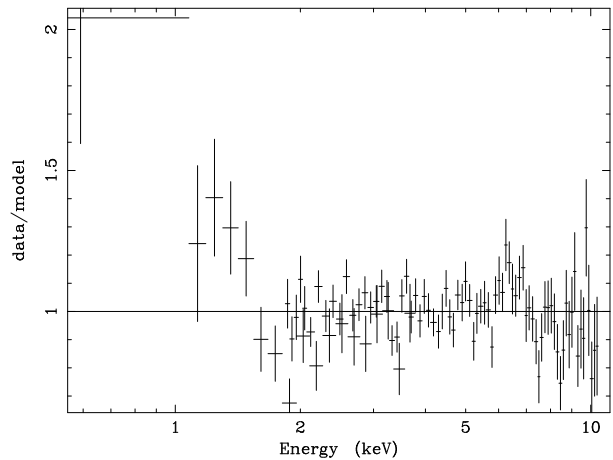}
\caption{Data to model ratio obtained from fitting the LECS and MECS
spectra with a simple absorbed powew law.}
\end{figure}
Due to the low statistics available below 2 keV, no deep investigation
is possible of the soft excess. We have tried to fit it with the
following models: (a) power law, (b) partial covering power law 
with photon index equal to the one characterizing the primary continuum
and (c)
Raymond--Smith plasma.

In each case, only galactic absorption was
assumed for this component. Each model provides a significant improvement 
in the quality of the fit ($>99\%$) and results 
in a photon index of 1.64$^{+1.30}_{-0.75}$ for model (a),
a column density of $\big(0.97^{+6.50}_{-0.42}\big)\times10^{22}$
cm$^{-2}$ 
with a covering fraction of 0.97$^{+0.03}_{-0.10}$ for model (b) 
and a temperature of 0.17$^{+0.14}_{-0.06}$ keV for model (c).
The ratio/data model for case (a) is shown in Fig. 2.

In all cases the 0.1--2.4 keV flux is $0.16\times10^{-11}$ erg cm$^{-2}$
 s$^{-1}$ in agreement with \emph{ROSAT} measurement. 
Despite the improvement in the quality 
of the fit, a local wiggle still remains around 1--2 keV in model (a) and (c) 
(see Fig. 2).
This may be due
to unresolved lines in this region or to an extra component (which requires
absorption in excess of the galactic value) not accounted for in the
fit.

\begin{figure}
\psfig{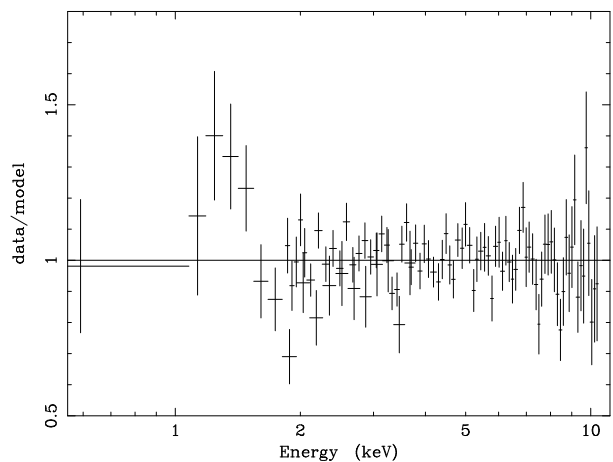}
\caption{Data to model ratio obtained from fitting the LECS and MECS
spectra with a power law, to account for the soft excess, and an absorbed
power law plus a gaussian narrow line to describe the high energy data.}
\end{figure}
Note that a similar soft X--ray spectral complexity of uncertain origin
was
also found by Turner et al. (1998) in their analysis of the \emph{ASCA}
data. 
Unfortunately the statistics in the low energy band prevents us 
to make any definite statement on the nature of the soft excess.
In any case the overall  LECS--MECS data on NGC 526A confirm basically the
previous 
results obtained by \emph{Ginga} and \emph{ASCA}, but allowing a spectral
characterization
during a different flux level: contrary to other Seyfert 2 galaxies, 
(Dadina et al. 2000; Turner et al. 1997) NGC 526A
seems to have a stable spectral behaviour despite a significant change
in the level of the continuum emission.
\subsection{The PDS high-energy spectrum}
The high energy spectrum (13--150 keV) measured by the PDS is well fitted
($\chi^2/\nu=10.9/12$) by a simple power law model with
$\Gamma=1.79^{+0.18}_{-0.20}$.

In the effort of verifying and quantifying the significance of this
spectral steepening above 10 keV, a broken power law (plus the narrow iron 
line)
model was used to describe the MECS--PDS data (1.8--150 keV). 
For simplicity the knee of the broken
power law was frozen at 10 keV, but a knee with free energy did not 
affected the results significantly.
The model provides a good ($\chi^2/\nu=89.5/84$) fit to the data and
the following photon indices: $\Gamma_{\rm 2-10}=1.56\pm0.07$ and
$\Gamma_{\rm 13-150}=1.78\pm0.2$.

It is still possible, however, for the source to have an intrinsically
flat spectrum. The observed spectral steepening at $E>13$ keV
is then due to the exponential cut--off of the primary power law.
We have tested 
this model against our data and found that the fit
($\chi^2/\nu=88.7/84$) is equally good (if not better) and  
gives a photon index: $\Gamma=1.52\pm0.08$ with a cut--off value at 
149$^{+635}_{-74}$ keV.

We cannot therefore discriminate between this model and a
broken power law one, but can only conclude at this stage that
the PDS spectrum is steeper than the MECS one and look for other
characterizations of the source to determine the real shape of the
primary continuum.  
\subsection{The iron line}
Although we do not find a broad line, inspection of Fig. 2 indicates
that 
some residuals are still present around 7 keV.
This combined with the slightly higher line energy measured (compared
with the expected 6.4 keV) suggests the presence of a more complex line
structure as also reported for the \emph{ASCA} data (Turner et al. 1998;
Weaver et al. 1998).
 
If the observed iron line is produced in the molecular torus (via
transmission and/or reflection), we expect an $EW\leq20$ eV for a
column of $\sim$ $10^{22}$ cm$^{-2}$ (Ghisellini, Haardt $\&$ Matt 1994);
we
therefore conclude that the observed line cannot be explained only by
transmission in the absorbing material.
Instead, the observed structure is reminiscent of a
relativistic  disk
line profile (George $\&$ Fabian 1991).

In view of this indication, we have fitted the data using the
\emph{diskline} model in {\sc XSPEC} (Fabian et al. 1989), assuming
a Schwarzschild 
black hole having inner radius $R_{in}=6r_{g}$, outer radius
$R_{out}=1000r_{g}$,  
(where $r_{\rm g}=MG/c^2$ is the gravitational radius and $M$ is the
mass of the black hole), and emissivity slope $\beta$~=~--2.5;
furthermore,
we fixed the line energy $E_{(\rm K \alpha) \rm D}$ at 6.4 keV, leaving
as free parameters 
the disk inclination ($i$) and the line normalization.

In the following, the soft--excess is always parameterized with a
scattering 
model (i.e. with a power law having the same photon index of the primary 
continuum) unless otherwise stated.
The \emph{diskline} model yields an improvement in the fit of
$\Delta\chi^{2} =4.2$ (for an equal number of degrees of freedom) with
respect to a narrow
line model, and provides an inclination angle $i$ and
an $EW_{\rm D}$ in agreement with the \emph{ASCA} results reported
by Weaver et al. (1997) (see Table 1).

We have also tried to
add a narrow line at 6.4 keV to the disk line component to account for the
presence of a torus, but found no statistically significant improvement
in the quality of the fit; in any case the torus component produces an
$EW< 98$ eV.
                              
Alternatively, the observed residual can be due to an extra narrow line
at an energy $E_{(\rm K \alpha)2}=6.9$ keV (see Table 1): the fit 
is equally good ($\chi^2/\nu=139.6/118$), if not
better, but the origin of this extra line is more difficult to account for
in a Compton thin source like NGC 526A.

Figure 2 also shows evidence for two possible edge structures
in the 7--9 keV range: an edge at 7.4 having an optical depth of 0.45
was indeed reported by Turner et al. (1997).

We have tried to add
such features in our spectrum but obtain no improvement in the fit.
The upper limit we found for the optical depth is 0.16 for an edge at
7.1 keV,
or 0.14 for an edge in the range 7.4--8.5 keV.
These limits are compatible with column densities
lower than 10$^{23}$ cm$^{-2}$, i.e. there is no strong requirement for
more absorption than observed.

However, in view of previous results on some flat spectra Seyfert 2
galaxies (Malaguti et al. 1999; Vignali et al. 1998), we have also tested
our data against a complex absorber model. This can be parameterized by 
a dual absorber model, where two columns cover both uniformly the source
(in this case the relative normalization gives the percentage of the
source
covered by each column), or by a partial plus uniform absorber model where
one absorber covers only partially the source, while the other covers it
totally. These two models are algebraically identical, but require
slightly different assumptions on the source geometry (see Cappi 1998).

We have tested both models against our \emph{Beppo}SAX broad data, and
obtained
a better fit ($\chi^{2}/\nu=136.7/117$) for the first model. Note that
in both cases no extra component is required to described the
\emph{soft--excess}, but if this is modeled with a scattering component,
the results do not change significantly.
The data indicate that 53$\%$ of the source is
covered by a column of $\sim$ 10$^{22}$ cm$^{-2}$, and 47$\%$ by a column
of a few 10$^{22}$ cm$^{-2}$ (see Table 1).

However, the spectrum remains flat since the data do not required a strong
additional absorber. Also the column densities observed can only explain a
line $EW$ of $\sim$ 50 eV, much lower than the value reported for this
model.
%
%
%
%
\begin{figure}
\psfig{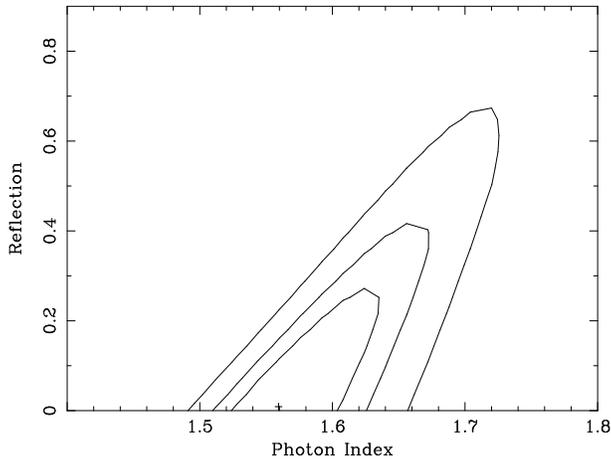}
\caption{Contour plot of the relative normalization between the primary
and Compton reflected spectral components versus the intrinsic photon
index.}
\end{figure}
\section{Defining the source spectral components}
The previous analysis already gives a number of useful indications.
First of all, the presence of complex absorption  
can not be used to explain the flat spectrum observed in the source,
nor the line $EW$ obtained by the data.
Second, the line parameterization indicates the presence of a reflection
component most likely due to an accretion disk.

We therefore conclude that reflection may play a role
in this source and may be responsible for the flat spectrum observed.

Therefore, we have tried to fit the data with the {\sc pexrav} model in
XSPEC 
(Magdziarz $\&$ Zdziarski 1995), which consists of an exponentially 
cut--off power law reflected from a neutral slab, plus
a narrow gaussian line;  
all these components are absorbed by intrinsic absorption as expected
in the case of reflection by an accretion disk.

Furthermore, we have fixed the 
disk inclination to 42$^{\circ}$ as estimated by the line spectral
analysis.
First, we assumed the cut--off to be high enough (10$^{4}$ keV) not to be 
effective in the fit: we obtain a photon index $\Gamma$ of $\sim$~1.6
and an upper limit to the reflection coefficient $R$ (the solid angle in 
units of 2$\pi$ subtended by the
reflecting matter) of $\sim$~0.3 (see Table 1).
The contour plot of $\Gamma$ versus $R$ (Fig. 3) clearly shows that in any
case the spectrum remains flat and  
incompatible with the canonical value of 1.9 even if the reflection is
at the maximum value allowed by the fit (i.e. $R<0.4$ at 90$\%$
confidence level).
This strongly suggests that the data prefer a flat spectrum and a small
reflection. 
If the cut--off energy is allowed to vary, we
obtain a best fit model (Fig. 4) with a cut--off at an energy $E_{\rm
C}$ at around 100 keV and $R<0.7$ (see Table 1); in this
case the improvement in the fit with respect to a
pure power law is $\sim$~95$\%$ or $\Delta\chi^{2}=7.2$
for two degrees of freedom.
%
%
%
%
\begin{table*}[t]
\caption{Principal models used to fit the data in the 0.1--150 keV band
(see text).}
\small
\smallskip
\begin{center}
\begin{tabular}{cccccc}
\hline
\hline
    &   &   &   &    \\  
Parameter&Double gaussian line&diskline&Reflection&Reflection&Dual
absorber\\
    &      &        & $E_{\rm C}=10^{4}$ keV& $E_{\rm C}$~free & \\
 &  & &    &  &  \\
\hline
\hline
    &  &  &    &  &    \\
$N_{\rm H}~(10^{22}$ 
cm$^{-2}$)&1.71$^{+0.23}_{-0.17}$&1.68$^{+0.23}_{-0.18}$&
1.72$^{+0.22}_{-0.27}$&1.56$^{+0.30}_{-0.20}$&3.17$^{+2.66}_{-1.32}$\\
   &   &   &   &  &    \\
$N_{\rm H_{\rm dabs}}~(10^{22}$
cm$^{-2}$)& - & -& -& -& 0.96$^{+0.27}_{-0.32}$ \\
    & &   &    &   &   \\
$\Gamma$ &1.56$^{+0.07}_{-0.03}$&1.57$\pm$0.04&
1.56$^{+0.07}_{-0.04}$&1.47$^{+0.13}_{-0.07}$ & 1.65$\pm$0.08\\
   &    &    &    &    & \\
$E_{\rm K \alpha}$~(keV) &6.4$^{(f)}$& - &6.53$^{+0.17}_{-0.13}$
&6.52$^{+0.20}_{-0.14}$& 6.54$^{+0.20}_{-0.14}$ \\
    &    &     &    &     \\
$EW$~(eV) &103$\pm$48& -&133$\pm$47&121$\pm$50 & 265$\pm$100\\
    &    &     &    &   &  \\
$E_{(\rm K \alpha)2}$~(keV) &6.9$^{(f)}$&-&  -  & - & -  \\
   &   &   &    &    &   \\
$EW_{2}$~(eV)&91$^{+53}_{-51}$&-& -   &  - &   -   \\
     &    &    &    &  &    \\
$E_{(\rm K \alpha) \rm D}$~(keV) &-&6.4$^{(f)}$ &  -  &   - &  - \\
    &   &     &     &   &    \\
$EW_{\rm D}$~(eV)   & -   & 283$^{+104}_{-94}$&  -   &  -  &   -  \\
    &    &    &    &   &    \\
$i$ (degrees)& - & 42.2$^{+7.1}_{-9.7}$& - & - &   - \\
    &    &     &    &   &    \\
$E_{\rm C}$~(keV)&-&-&10$^{4}$$^{(f)}$ &112$^{+234}_{-50}$ & -    \\
     &     &     &     &   &    \\
$R$&-& -& $<$~0.3& $<$~0.7$$&   -    \\
    &    &    &    &   &   \\ 
$\chi^{2}/\nu$  &139.6/118 &141.1/118&145/117&138.1/116 &  136.7/117\\
       &    &    &    &   &    \\
\hline
\hline
\end{tabular}
\begin{list}{}{}
\item[$^{(f)}$] the parameter is fixed at the value reported in the
table.
\end{list}
\end{center}
\end{table*}

Comparison with the previous model suggests that the data are more
sensitive to the introduction of the cut--off energy rather than to the
reflection component.

%
%
%
%
%
\begin{figure}
\psfig{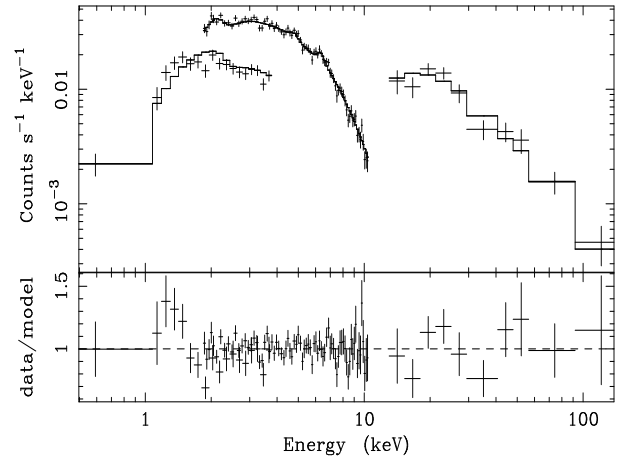}
\caption{MECS and PDS spectra best--fit model (\emph{upper panel}) and 
data to model ratio (\emph{lower panel})
for the {\sc pexrav} model
(in which the cut--off energy is allowed to vary) plus a narrow gaussian
line.}
\end{figure}
\begin{figure}
\centerline{\psfig{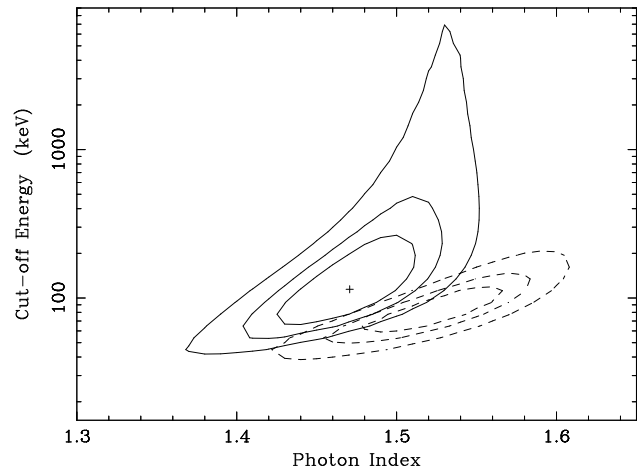}}
\vspace{5mm}
\caption{Contour plot of the cut--off energy of the primary power law   
component versus the photon index for two possible reflection values:
$R=0$ (solid line) and $R=0.7$ (dashed line).}
\end{figure}
In order to better define the cut--off energy, we have fixed the line
parameters and the intrinsic absorption to their best values for
two possible reflection value within the estimated range of
uncertainty ($R=0$ and $R=0.7$): the
contour plot of $\Gamma$ versus cut--off energy for these cases is shown
in Fig. 5 which clearly localize the power law exponential decline
in the range 40--400 keV.

We have also tried an ionized reflection model, which provides a slightly 
higher reflection component ($R$ $\sim$ 0.4), an ionization parameter
$\chi_{i}$
of $\sim$~5 and a cut--off at around 140 keV ($\chi^{2}/\nu=137.4/115$).
However, even in this case the spectrum remains flat as the reflection is
insufficient to steepen the high energy continuum.

Although the dual absorber model is statistically a slightly better fit
with respect to the reflection model, it has more difficulties in
explaining
the iron line, both in terms of $EW$ value and spectral profile.

In view of this, we assume valid the reflection model and try, in the
following, to verify the self--consistency of this assumption.
\section{Discussion}
The overall picture that emerges from the NGC 526A \emph{Beppo}SAX data is
that of a
flat spectrum source absorbed by a column density
of $\sim$~10$^{22}$ cm$^{-2}$. In the context of the reflection
model,
the source has a small reflection component ($R\le0.7$) and an iron line
with an $EW$ of 121 eV. The reflection component originates 
in an accretion disk seen at an angle of $\sim$~42$^\circ$.

Recently Zdziarski et al. (1999) found a strong correlation between the   
amount of Compton reflection from neutral gas and the intrinsic
spectral
slope, in the sense that  the reflection strength decreases  with
decreasing $\Gamma$. This correlation has a simple physical interpretation
as long as the same medium that is responsible for the observed Compton  
reflection also plays a role in the cooling of the source hot plasma.
Less reflection provides less cooling which in turn produces a flatter
spectrum (Zdziarski, Lubi\'nski $\&$ Smith 1999). As the line is also
produced in the same medium, a
correlation between the strength
of the reflection and the width of the Fe$_{\rm K \alpha}$ is also
expected (Lubi\'nski $\&$ Zdziarski 2000).
%
%
%
%
\begin{table*}[t]
\caption{Main characteristics of the sources found in the 2MASS
Atlas Images.}
\small
\smallskip
\begin{center}
\begin{tabular}{cccccccc}
\hline
\hline
    &   &   &   &   &   &   &      \\
Source&$\alpha$(2000)&$\delta$(2000)&m$_{\rm b}$&m$_{\rm r}$& J &H & K\\
 &  &   &    &   &   &     &     \\
\hline
\hline
    &  &  &    &  &    &    &        \\
1 &01$^{h}$22$^{m}$47.6$^{s}$&--34$^{\circ}$48$'$46.6$^{\prime
\prime}$&18.4&17.1&16.165&15.543&14.650\\
    &     &     &    &    &    &    &    \\
2&01$^{h}$22$^{m}$48.2$^{s}$&--34$^{\circ}$48$'$57.3$^{\prime
\prime}$&20.1&17.9&16.345&15.400&15.088\\
   &   &    &    &    &    &    &\\
\hline
\hline
\end{tabular}
\end{center}
\end{table*}
%
%
%
%
%
%
%

For a power law of index
$\Gamma$~$\sim$~1.6 and a viewing angle of 42$^\circ$ the expected
relationship between $EW$ and $R$ is $EW$~$\sim$~(140--160)~$\times$~$R$ 
(George $\&$ Fabian 1991; \.Zycky $\&$ Czerny 1994).
In our case this yields an $EW<100$ eV if $R<0.7$,
in agreement with the uncertainty range obtained with the reflection 
model, especially if part of the $EW$ (at least 20 eV) is due to
transmission in the absorbing medium and the reflecting medium is  
mildly ionized.
Therefore, the line $EW$ value and
spectral profile are well described by a mildly ionized medium.
Also our observation of a cut--off at around 100 keV (albeit the
uncertainty range is large) agrees with the observed tendency for
flat sources to avoid high values of the cut--off energy.
%
%
%
%
%
%

\vspace{-1.5mm}
In a recent analysis
of the cut--off versus power law index of a sample of Seyfert 1 galaxies
observed with \emph{Beppo}SAX, Matt(2000) finds that sources as flat as
NGC 526A
have a cut--off below 200 keV as indeed observed here. Although it is
still debatable if  this has a
physical meaning or is due
to a selection effect (i.e it is more difficult to determine an
high energy cut--off in steep sources), nevertheless our observation
confirms
this trend. The presence of a low energy cut--off also explains why the
PDS data are steeper than the MECS ones.

Finally, this picture also explains the stable spectral shape observed
in NGC 526A despite large changes in its flux level. If reflection is
small at all flux levels, the power law component always dominates; 
therefore no spectral variations are observed as the source luminosity
varies unless the primary power law shape changes.

On the basis of the above considerations, we are led to conclude that
the
overall picture is consistent with NGC 526A being an intrinsically flat
spectrum source.
On the other hand, NGC 526A is a bona--fide Seyfert 2 galaxy, as discussed
in the introduction, and it is therefore a crucial object for testing the
Unified Model.
So far NGC 526A is the only Seyfert 2 confirmed as a flat spectrum source,
the other objects reported have been found to be compatible with 
the "canonical" Seyfert spectral index of 1.8--2.0, when more complex
models are considered.                                                         %

In particular, analysis of \emph{Beppo}SAX data of
other sources from the Smith $\&$ Done sample, namely NGC 4507, NGC 2110
and MCG--5--23--16 (Lanzi 2000; 
Malaguti et al. 1999; 
Malizia et al. 2001) are all compatible with a canonical spectrum when
analyzed over a broad band. 
Thus our observation suggests that flat Seyfert 2 galaxies exists but are
not
common. This finding does not necessarily contradicts the Unified Models,
since flat Seyfert 1 galaxies are also observed: Matt (2000) lists 3 such
sources 
in his sample of Seyfert 1 galaxies, the most prominent being the atypical
NGC 4151.

The X--ray continuum of Seyfert galaxies is commonly believed to be
produced by Compton scattering of soft photons on a population of hot 
electrons. Assuming thermal equilibrium and a plane parallel geometry
for the Comptonizing plasma above the accretion disk, the average
properties of Seyfert galaxies could be naturally accounted for 
(Haardt $\&$ Maraschi 1991).

%
%
\begin{figure}[t]
\psfig{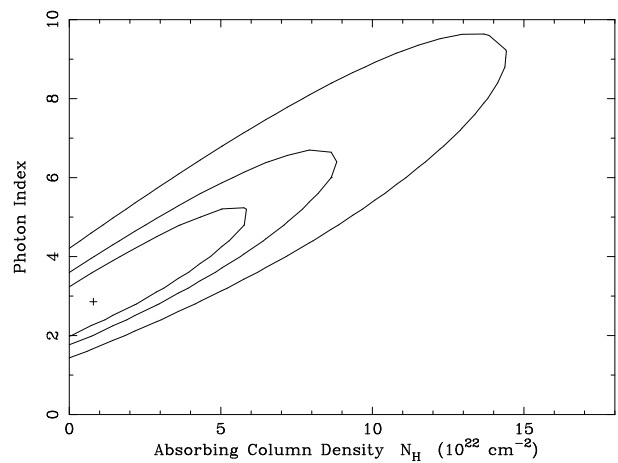}
\caption{Confidence contours for the photon index versus the equivalent
Hydrogen column density in the absorbed power law model obtained for
the source observed within the MECS field of view.}
\end{figure}
However, this model is unable to easily 
explain spectra flatter than $\Gamma$ $\sim$~2. This has been
traditionally
the main concern about the existence of flat spectra source.

However Petrucci et al. (2000) have shown that the cut--off power
law continuum + reflection usually employed to fit X--ray data is a too
simple model as it
derives a lower temperature and an higher optical depth for the hot
plasma than obtained with accurate thermal Comptonization models,
which treat carefully the anisotropy effects in Compton processes.
In these models, 
the LECS and MECS data determine the slope below the anisotropy break.
Above this break the intrinsic spectrum is steeper. It can thus fit the
PDS data without an additional steepening beyond 100 keV. 
Within these models therefore flat spectra sources can be easily 
accounted for, thus 
removing the uncomfortable situation that they are "strange/peculiar"
objects to our global understanding of Seyfert galaxies.
%
%
%
%
%
%
\section{Appendix}
In this section we briefly report some characteristics of the
other source detected in the MECS field of view.

Within the MECS error box, we find a faint ROSAT source at position 
$\alpha(2000)=01^{h}22^{m}50.3^{s}$,
$\delta(2000)$~=~--34$^{\circ}47'54^{\prime \prime}$ (37 arcsec positional
uncertainty), having a flux, in the 0.1--2.4 keV band,
of $\sim$ $4\times10^{-13}$ erg cm$^{-2}$ s$^{-1}$. Also in the 
MECS error box of 1$'$, we find two sources from the MASS 2nd 
Incremental Release Point Source Catalogue, whose characteristics 
are shown in Table 2. The first object is identified with the galaxy
APMUKS(BJ) B012030.16--350426.7. Neither of the 2$\mu$m sources
fall within the ROSAT error box.
   
Notwithstanding the low statistics of the data, we tried to analyze the
MECS source spectrum extracted from a region centered in the 
source with a radius of 2$'$.

First, we fitted the data with a simple power law: the model is
satisfactory ($\chi^{2}=20.2/17$) and yields a photon index
$\Gamma=2.55^{+0.64}_{-0.77}$.
The addition of intrinsic
absorption to the previous model does not improve the quality of the
fit ($\chi^{2}=20/16$) and gives a 90$\%$ upper limit of
$6.5\times10^{22}$ cm$^{-2}$. 
The confidence contours of the photon
index against the absorbing column density are shown in Fig. 6.

The flux in the 2--10 keV band turned out 
to be $3.4\times10^{-13}$ erg cm$^{-2}$ s$^{-1}$.
\begin{acknowledgements}
This research has made use of SAXDAS linearized and cleaned event files produced at
the \emph{Beppo}SAX Science Data Centre. G.M., G.G.C.P., M.C. and
L.B. acknowledge 
financial support from the Italian Space Agency. We would like to thank 
the referee Dr. H. Inoue for the very useful comments which have improved
the quality of this work.
\end{acknowledgements}
%


\begin{thebibliography}{}
   \bibitem[1991]{Aw91} Awaki, H., Koyama, K., et al. 1991, PASJ, 43, 195
   \bibitem[1997]{Boella97a} Boella, G., et al. 1997a, A\&AS, 122, 299
   \bibitem[1997]{Boella97b} Boella, G., et al. 1997b, A\&AS, 122, 327
   \bibitem[1996]{Cap96} Cappi, M., et al. 1996, ApJ, 456, 141
   \bibitem[1998]{Cappi} Cappi, M. 1998, PhD thesis, University of
Saitama  
   \bibitem[1989]{Fab89} Fabian, A.C., et al. 1989, MNRAS, 238, 729
   \bibitem[1999]{SAXabc99} Fiore, F., Guainazzi, M., Grandi, P. 1999,
SAXabc, vs. 1.2, Cookbook for \emph{Beppo}SAX NFI Spectral Analysis.  
   \bibitem[1997]{Fro97} Frontera, F., et al. 1997 A\&AS, 122, 357
   \bibitem[1991]{GandF91} George, I.M., Fabian A.C. 1991, MNRAS, 249, 352
   \bibitem[1994]{Ghis94} Ghisellini, G., et al. 1994, MNRAS, 267, 743  
   \bibitem[1979]{Grif79} Griffiths, R.E., et al. 1979, ApJ, 230, L21
   \bibitem[2000]{Lanzi00} Lanzi, R. 2000, Master thesis, University of
Bologna
   \bibitem[2000]{Lub00} Lubi\'nski, P., Zdziarski, A. 2001, MNRAS,
323, L37
   \bibitem[1995]{MandZ} Magdziarz, P., Zdziarski, A.A. 1995, MNRAS,
273, 837  
   \bibitem[1999]{Malag99} Malaguti, G., et al. 1999, A$\&$A, 342, L41
   \bibitem[1999]{Mali99} Malizia, A., Bassani, L., et al. 1999, ApJ,
519, 637
   \bibitem[2000]{MAng00} Malizia, A., et al. 2001, in preparation
   \bibitem[2000]{Matt00} Matt, G. 2001, ApJ, in press 
(astro--ph/0007105)
   \bibitem[1997]{Parm97} Parmar, A.N., et al. 1997, A\&AS, 122, 309
   \bibitem[2000]{Petr00} Petrucci, P.O., et al. 2001, ApJ, in press 
(astro--ph/0004118)
   \bibitem[1996]{Poll96} Polletta, M., Bassani, L., et al. 1996 
ApJS, 106, 399
   \bibitem[1996]{Rush96} Rush, B., et al. 1996, ApJ, 471, 190
   \bibitem[1996]{SmithDone96} Smith, D. A., and Done, C. 1996, MNRAS, 280, 355
   \bibitem[1997]{Turner97} Turner, T.J., George, I.M., Nandra, K., et al. 
~1997, ApJS, 113, 23
   \bibitem[1998]{Turner98} Turner, T.J., George, I.M., Nandra, K., et al.
~1998, ApJ, 493, 91
   \bibitem[1998]{Vig98} Vignali, C., et al. 1998, A\&A, 333, 411
   \bibitem[2000]{Vog00} Voges, W., et al. 2000, IAUC, 7432, 3
   \bibitem[1985]{W98} Weaver, K.A., Reynolds, C.S. 1998, ApJ, 503, L39
   \bibitem[1992]{Win92} Winkler, H. 1992, MNRAS, 257, 677
   \bibitem[1999]{Zdi99} Zdziarski, A.A., Lubi\'nski, P., $\&$ Smith, D.A.
1999 MNRAS, 303, L11
   \bibitem[2000]{Z2000} Zdziarski, A.A., et al. 2000, ApJ, 542, 703
   \bibitem[1994]{ZandC} \.Zycki, P.T., Czerny, B. 1994, MNRAS, 266, 653
\end{thebibliography}
\end{document}